\documentclass[aps,prl,preprint,showpacs,,amsmath,amssymb]{revtex4}
\usepackage{graphicx}
\usepackage{bm}

\begin{document}

\title{Roton dipole moment}

\author{V. P. Mineev}

\affiliation{ Commissariat \`a l'Energie Atomique, INAC/SPSMS, 38054
Grenoble, France}
\date{\today}

\begin{abstract} 
The roton excitation in the superfluid $^4He$ does not possess a stationary dipole moment.
However, a roton has an instantaneous dipole moment, such that at 
any given moment one can find it in the state either with positive or with negative dipole moment projection 
on its momentum direction.
The instantaneous value of electric dipole moment of roton excitation  is evaluated. The result is in reasonable agreement with recent experimental observation of 
the splitting of microwave resonance absorption line at roton frequency under  external electric field.
\end{abstract}


\maketitle

\bigskip
The interaction of  electromagnetic radiation  with superfluid $^4He$ was the subject of recent intensive investigations undertaken by Kharkov experimental group.
Among several new effects there was observed the 
resonance absorption of microwaves at the frequency corresponding to  the roton gap $\Delta(T)$ of the phonon-roton excitation spectrum.\cite{Rybalko2007}  As well the inverse effect of generation of electromagnetic radiation with frequency of the roton gap by heat flow in superfluid helium has been detected.\cite{Ryb1,Ryb2}
The photon momentum $\hbar {\bf k}_{ph}$ is many orders less than the roton momentum $\hbar {\bf k}_r$, hence, 
the creation of a single roton  by means of single photon absorption is prohibited due to momentum conservation law. 
This problem can be, however,  passed round  noting  that 
roton momentum is compensated by momentum of liquid flow\cite{footnote} ${\bf P}_l$ arising in some macroscopic volume $V$,  which is much larger than the volume per atom in liquid helium  
\begin{equation}
{\bf P}_l=Vmn{\bf v}=\hbar{\bf k}_{ph}-\hbar{\bf k}_r\approx-\hbar{\bf k}_r.
\end{equation}
Here m is the mass of $^4He$ atom and $n$ is the fluid density 
The process can happen  practically without  a change of flow energy 
 \begin{equation}
  E_l=\hbar ck_{ph}-\Delta=Vn\frac{mv^2}{2}=\frac{P_l^2}{2Vnm}\approx\frac{\hbar^2k_r^2}{2Vnm}\ll\Delta.
\end{equation}  
An example of such a kind transition between the two states of superflow with the same energy but different momenta was found by Volovik \cite{Volovik} who considered 
quantum mechanical formation of vortices from 
the homogeneosly  moving superfluid. This case, however, the momentum of the liquid is not conserved because of violation of translational symmetry by inhomogeneity - a hemisphere on the wall of container with liquid helium. The same is probably happens in the experiments under discussion: the inhomogeneities on the walls of the container are responsible for the non-conservation of momentum. They provide the necessary matrix element for the transition between the states with different momentum of liquid.\cite{6}

 The resonance microwave absorption in liquid $^4He$ at the frequency corresponding to the roton minimum 
 can be interpreted as the  evidence of an electric dipole moment of roton excitations. Indeed, the following investigations have demonstrated that the resonance absorption line at roton frequency splits 
on two lines by the constant electric field.\cite{Rybalko2009}  The splitting at small enough fields depends linearly on the
field value that corresponds to the roton dipole moment $d\approx 10^{-22} ~CGSE~ units$.

The existence of the roton dipole moment  is unnatural  from the symmetry point of view.  A roton is collective excitation that is described by a compact in space wave function with typical size about few  interatomic distances.\cite{Feynman}  A roton possess the definite  momentum characterized by its modulus and direction. This causes the local space parity violation inside the region occupied by the roton wave pocket. On the other hand the  state with definite momentum is characterized by the time 
reversal breaking such that only the product space and time inversion $PT$ is conserved quantity.
On the contrary the polar vector of the dipole moment  changes its sign under space inversion and it is not changed under the time reversal. So, the roton  and the dipole obey the different symmetry. Hence, the roton cannot possess   the stationary dipole moment. We shall demonstrate, however, that the roton dipole moment can be treated as a sort of nonstationary phenomenon.

 According to the Feynman \cite{Feynman} conjecture each roton is  described
by many particle wave function corresponding to dipole distribution of velocity  field of $^4He$ atoms.
Roughly speaking, roton is similar to "a vortex ring of such small radius that only one atom can pass through the center.\cite{9} Outside the ring there is a slow drift of atoms returning for another passage through the ring."  The roton  momentum is approximately equal to the inverse interatomic distance $\hbar k_r\approx \hbar/a$.   An atom passing through the ring center, first acquires this momentum, then it slows down its motion to go around and come back to the initial point where it is accelerated again.  The corresponding force which is necessary to get and then to lose such a momentum is 
\begin{equation}
{\bf f}(t)\approx \frac{\hbar^2g(t)}{ma^3}~\hat{\bf k}_r,
\end{equation} 
where $g(t)=\sum_{n\ge1}c_n\sin n\omega t$ is a periodic function of time with period $\tau=2\pi/\omega\approx ma^2/\hbar$, and $\hat{\bf k}_r$ is the roton momentum direction.
This force acting on given atom from the side of other atoms pushes it through the ring center. 
 
The helium atom in the ground state does not have an electric dipole moment.  The uneven motion  of helium atom under the action of force given by eqn.(3)  causes the deformation of the atom electronic shell. 
To estimate  the dipole moment caused by this deformation let us write the Hamiltonian
of two electrons in He atom as
\begin{equation}
\hat H=\hat H_0+{\bf F}({\bf r}_1+{\bf r}_2),~~~~~~{\bf F}={\bf F}(t)=e{\bf E}+ {\bf f}(t),
\end {equation}
where the first term $\hat H_0$ presents the  unpertubed electron Hamiltonian of helium atom and the second term is a
potential of perturbation determined by the external electric field and by the force ${\bf f}(t)$ introduced above. The  vectors   ${\bf r}_1$ and ${\bf r}_2$ gives positions of two electrons relative  to the nuclei position.   The Hamiltonian  properly describes the electronic state of helium atom in usual approximation
when the electron mass is much smaller than the mass of atom $m_e\ll m$. The dipole moment is given by  the average of $e({\bf r}_1+{\bf r}_2)$ over the the ground state wave function.
The motion of the atom nuclei is described by separate equation and has no influence on the dipole moment value. 

The frequency of perturbation $\omega$ is much smaller than the distance between the energies of the ground  and the first excited state of $He$ atom: $\hbar\omega\ll (E_1-E_0)$.  Hence, one can prove  (see eg \cite{Landau}) that electrons  in this atom are in the quasi stationary state characterized by 
the wave function
\begin{equation}
\Psi_0^\prime(t)\approx\left [\Psi_0-\frac{\langle\Psi_0| {\bf F}({\bf r}_1+{\bf r}_2)|\Psi_1\rangle}{E_1-E_0}\Psi_1\right ] e^{-\frac{iE_0t}{\hbar}}
\end {equation}
Here, $\Psi_0$, $\Psi_1$ are the wave functions of the ground and the first excited states of He atom correspondingly. We neglect here admixture of the higher excited states.

The correction to the electron energy is
\begin{equation}
E^\prime_0-E_0 \approx -\frac{r_{at}^2F^2}{E_1-E_0}
\label{e3}
\end{equation}
Here, $r_{at}$ is the size of electron wave function of helium atom ( hard core radius in the potential of interaction between two He atoms). 

The linear in respect of electric field ${\bf E}$ term in (\ref{e3}) gives roton dipole moment
\begin{equation}
{\bf d}\approx\frac{2er_{at}^2}{E_1-E_0}~{\bf f}.
\end{equation}
This value obeys the proper symmetry properties:  being odd in respect of space inversion it is even in respect of time reversal.
Its projection on momentum direction is
\begin{equation}
d\approx\frac{2e\hbar^2r_{at}^2}{ma^3(E_1-E_0)}g(t).
\end{equation}
Hence, as it was expected, the time average of the roton dipole moment is equal to zero.
However, a roton  possess an instantaneous dipole moment, such that at any  given moment one can find it in the state either with positive or with negative projection of dipole moment on its momentum direction.
To get the correspondence with the experimental observations one should assume that the time of transition between these two states is much shorter than period $\tau$ but still much longer than $\hbar/(E_1-E_0)$.  The latter condition provides the validity  of quasistationary approximation  has been used.

Substituting the numerical values and taking into account that $ r_{at}\approx a$ is about few Angstroms and $(E_1-E_0)\approx 20~eV$ we obtain $d\approx \pm 10^{-22} ~CGSE~ units$. This corresponds to the experimentally determined value of roton dipole moment. The roton dipole moment is temperature independent but increases with pressure.

The author is indebted to A. Rybalko and E. Rudavskii for the kind introduction to the vast field of their
experimental results. I am also grateful to L. Melnikovsky and G. Volovik for the useful and stimulating discussions.


\begin{thebibliography}{99.}
\bibitem{footnote}At  frequency of resonance absorption, which is much higher than the  temperature,  the hydrodynamic notions of the superfluid and the normal flows lose a sense.
\end{thebibliography}
\end{document}